# Understanding the growth mechanism of BaZrS$_3$ chalcogenide perovskite thin films from sulfurized oxide precursors

Santhanu Panikar Ramanandan,[1§] Andrea Giunto,[1§] Elias Z. Stutz,[1] Benoit Xavier Marie Reyner,[1] Iléane Tiphaine Françoise Marie Lefevre,[1] Marin Rusu,[2] Susan Schorr,[2,3] Thomas Unold,[2] Anna Fontcuberta i Morral,[1,4] José Márquez Prieto,[2,5]* Mirjana Dimitrievska,[1,6]*

1 - Laboratory of Semiconductor Materials, Institute of Materials, Faculty of Engineering, Ecole Polytechnique Fédérale de Lausanne (EPFL), 1015 Lausanne, Switzerland.

2 - Department of Structure and Dynamics of Energy Materials, Helmholtz-Zentrum Berlin für Materialien und Energie GmbH, Hahn-Meitner-Platz 1, 14109 Berlin, Germany

3 – Institute of Geological Sciences, Freie Universitaet Berlin, Maltese St. 74-100, 12249 Berlin, Germany

4 - Institute of Physics, Faculty of Basic Sciences, Ecole Polytechnique Fédérale de Lausanne, 1015 Lausanne, Switzerland.

5 - Humboldt University of Berlin, Unter den Linden 6, 10117 Berlin, Germany.

6 - Transport at Nanoscale Interfaces Laboratory, Swiss Federal Laboratories for Material Science and Technology (EMPA) Ueberlandstrasse 129, 8600 Duebendorf, Switzerland

§These authors have contributed equally to this work

*corresponding authors: mirjana.dimitrievska@empa.ch ; jose.marquez@physik.hu-berlin.de ;

## Abstract

Barium zirconium sulfide (BaZrS$_3$) is an earth-abundant and environmentally friendly chalcogenide perovskite with promising properties for various energy conversion applications. Recently, sulfurization of oxide precursors has been suggested as a viable solution for effective synthesis, especially from the perspective of circumventing the difficulty of handling alkali earth metals. In this work, we explore in detail the synthesis of BaZrS$_3$ from Ba-Zr-O oxide precursor films sulfurized at temperatures ranging from 700 °C to 1000 °C. We propose a formation mechanism of BaZrS$_3$ based on a two-step reaction involving an intermediate amorphization step of the BaZrO$_3$ crystalline phase. We show how the diffusion of sulfur (S) species in the film is the rate-limiting step of this reaction. The processing temperature plays a key role in determining the total fraction of conversion from oxide to sulfide phase at a constant flow rate of the sulfur-containing H$_2$S gas used as a reactant. Finally, we observe the formation of stoichiometric BaZrS$_3$ (1:1:3), even under Zr-rich precursor conditions, with the formation of ZrO$_2$ as a secondary phase. This marks BaZrS$_3$ quite unique among the other types of chalcogenides, such as chalcopyrites and kesterites, which can instead accommodate quite a large range of non-stoichiometric compositions. This work opens up a pathway for further optimization of the BaZrS$_3$ synthesis process, straightening the route towards future applications of this material.

## Introduction

Extensive harvesting of solar energy is required to minimize the use of fossil fuels for energy generation and to reduce $CO_2$ emissions. From the materials research perspective, new solutions based on abundant and non-toxic resources need to be developed for sustainable future deployment of solar energy on a large scale. With this in mind, chalcogenide perovskites are recently explored as a new wide-bandgap alternative for thin film photovoltaic (PV) absorbers.[1–3] They mostly contain Earth-abundant and non-toxic elements and they exhibit extraordinary chemical and thermal stability.[4]

Encouraging optoelectronic properties have been demonstrated experimentally, including an extraordinarily high absorption coefficient,[5,6] high luminescence efficiency,[7–10] relatively large charge carrier mobilities,[11] and the capability of being chemically doped to become both, *n* and *p*-type semiconductors.[7] These materials are predicted to be "defect tolerant" by ab-initio calculations: Detrimental defects with energy levels in the middle of the bandgap have high formation energies, thus being unlikely to exist in high concentrations in the material.[6]

From all the chalcogenide perovskites experimentally demonstrated, the most studied compound so far is $BaZrS_3$, which crystallizes in the $GdFeO_3$-type perovskite structure (space group Pnma).[12–15] $BaZrS_3$ has been reported to be a direct semiconductor with a bandgap value in the range of 1.8 to 2.0 eV,[5,8,16,17] making this compound attractive for a top cell in a photovoltaic tandem with Si, or for solar water-splitting applications. A variety of synthesis routes for thin films have recently appeared,[16–21] mostly based on physical vacuum deposition (PVD) methods. Most of these reports rely on a two-stage process in which an amorphous precursor film of Ba-Zr-O or Ba-Zr-S is first deposited, and then $BaZrS_3$ is crystallized in a second annealing step, often in presence of a reactive atmosphere containing sulfur.

Some of the elements constituting chalcogenide perovskites are extremely sensitive to air in their metallic form and as sulfide binaries. This is for example the case of metallic Ba or $BaS_2$. Using these materials as precursor films thus requires avoiding air exposure. Because of this inconvenience, many of the reported synthesis routes for $BaZrS_3$ use oxide precursor films. It has been suggested that this synthesis route has the additional advantage that the optoelectronic properties of the thin films can be controlled by the partial replacement of O by S in the film, ultimately allowing to tune the material bandgap.[22] However, this remains to be understood considering that as noted by Clearfield in powder samples,[14] and later verified by Marquez *et al.* in thin films,[17] $BaZr(S,O)_3$ does not form a crystalline solid solution.

The main disadvantage of using oxides as precursors is that the replacement of O by S to form $BaZrS_3$ is an energetically expensive process, requiring synthesis temperatures exceeding 800 °C.[1] This implies potential challenges for the future growth of these compounds in most of the commonly used conductive transparent substrates. At the moment, it is not clear whether the high-temperature requirement for crystallizing $BaZrS_3$ thin films is due to a limited diffusivity of the S atoms within the oxide precursor film, or limitations in the chemical reaction due to the formation of oxide intermediate compounds or secondary phases. Understanding the growth and formation mechanisms of chalcogenide perovskites will greatly accelerate the design of new synthesis routes for this new material class in thin films, enabling control of their composition and optoelectronic properties for their implementation in devices.

This work explores the formation mechanism involved in converting Ba-Zr-O precursor layers into $BaZrS_3$ by sulfurization at high temperatures between 700 °C and 1000 °C. Detailed morphological and compositional assessment of formed thin films was performed by (scanning) transmission electron microscopy (S/TEM) coupled with energy dispersive x-ray spectroscopy (EDX) and selective area electron diffraction (SAED) analysis on cross-sections of the thin films. Complimentary phase identification was done using grazing incidence wide angle X-ray scattering (GIWAXS) and Raman spectroscopy. Based on these results, we explore and discuss the efficiency of the synthesis reaction for converting Ba-Zr-oxides into Ba-Zr-sulfides, the possible limitation of S diffusion into the BaZrO$_3$ layers, and the preferential formation of secondary phases that could influence the formation of BaZrS$_3$.

## Experimental section

### Material preparation

This study uses samples from Ref. [[17]], where the synthesis procedure was described. A schematic illustration of the HZB preparation process is shown in **Figure 1**. A thin amorphous layer of Ba-Zr-O with a thickness of 150 nm was deposited by Pulsed Laser Deposition (PLD) on a quartz glass substrate at room temperature (**Figure 1a**) and then annealed under a continuous flow of 5% H$_2$S(g) in Ar at different temperatures ranging from 700 °C to 1000 °C for 30 minutes (**Figure 1b**). As shown in **Figure 1c**, this process resulted in a series of Ba-Zr-S-O thin films with a gradual colour change as a function of the sulfurization temperature, which correlates with the change in S composition. More information on the colouring, composition and optoelectronic characterization of the samples is reported in Ref. [17].

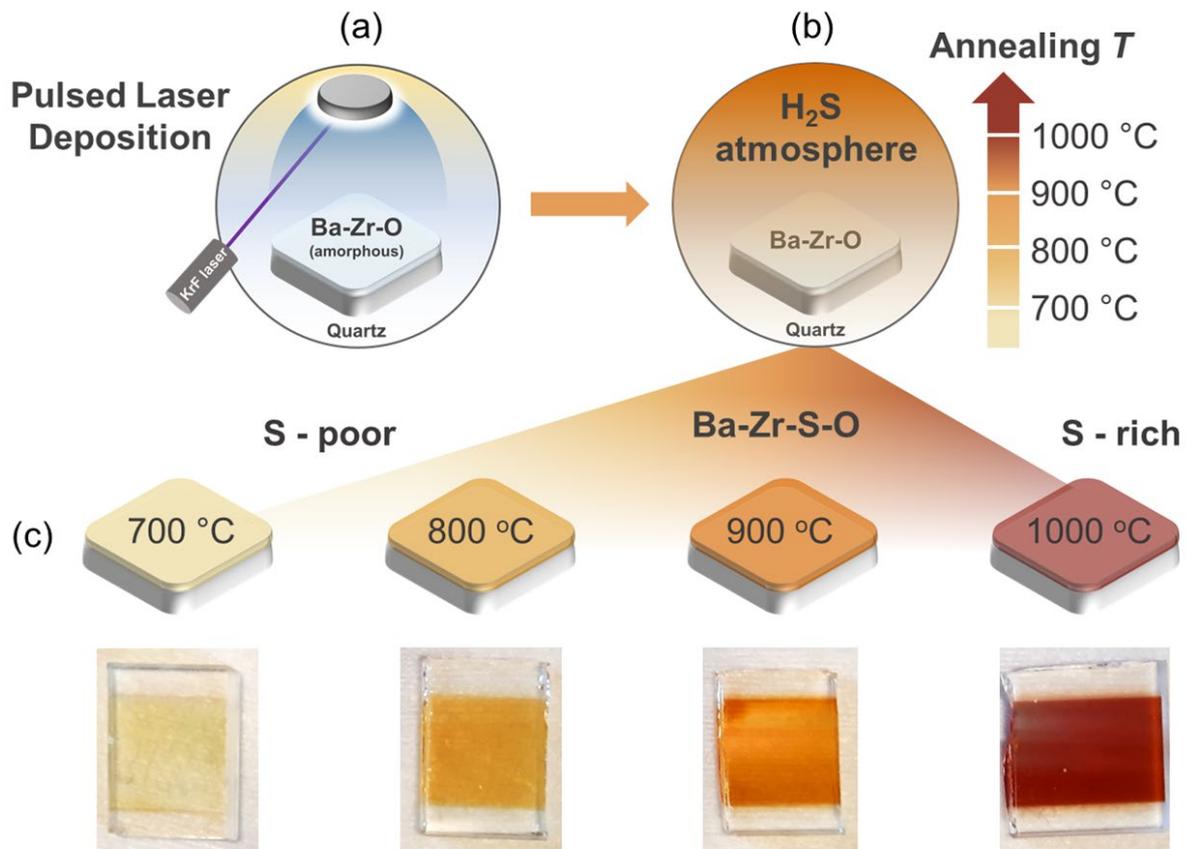

**Figure 1. Schematic illustration of the thin film synthesis procedure reported in Ref.** [17]: (a) A series of ~150-nm-thick amorphous Ba-Zr-O films on quartz glass substrates are synthesized by pulsed laser deposition (PLD) and (b) subsequently annealed under a continuous flow of 5% H$_2$S(g) in argon at various temperatures from 700 °C to 1000 °C. (c) The obtained Ba-Zr-S-O thin films show a gradual color change due to the variation in S composition. The color changes from a pale yellow for the S-poor sample sulfurized at 700 °C, to orange hues for samples with medium S compositions sulfurized at 800 °C and 900 °C, to finally dark red for S-rich sample sulfurized at 1000 °C.

## Characterization methods

**Scanning electron microscopy (SEM)**: Morphology of the samples' surface was characterized with a Zeiss Merlin SEM microscope operated at 3 kV and using in-lens detector.

**(Scanning) Transmission electron microscopy and energy-dispersive x-ray spectroscopy**: Cross-sections of the thin film samples were prepared with a dual-beam focused ion beam and scanning electron microscope (FIB-SEM, Zeiss Nvision 40). Annular dark field (ADF) STEM image and EDX elemental maps were collected using FEI Talos transmission electron microscope operating at 200 kV.

**Grazing Incidence Wide Angle X-ray Scattering:** GIWAXS measurements were performed on a Bruker Discover Plus equipped with a rotating anode and a Dectris Eiger2 500K detector operating in 2D mode. Collimating optics of 300 micron were used to select the beam shape. Grazing incident angles of 0.3º, 0.5º, 1º, 1.5º and 2º were used.

**Raman spectroscopy**: Raman measurements were performed complementary to the GIWAXS characterization to identify formed phases,[23] defects,[24–27] inhomogeneities,[28] and crystallinity.[29] Raman spectroscopy was implemented in backscattering configuration at 12 K. The 488 nm and 532 nm line of a Coherent sapphire optically pumped semiconductor lasers were used for excitation. The beam was focused on the sample with a microscope objective with a numerical aperture of 0.75, resulting in a 1 μm diameter spot, reaching a radiant power of the order of 500 μW. The backscattered light was analyzed using a TriVista triple spectrometer with 900 cm$^{-1}$, 900 cm$^{-1}$ and 1800 cm$^{-1}$ gratings in subtractive mode and a Princeton Instrument liquid nitrogen cooled multichannel CCD PyLoN camera. All spectra were calibrated based on the reference sulfur Raman spectrum.

**Rutherford backscattering spectrometry (RBS):** RBS measurements, which are performed by EAG Laboratories, were taken with a nearly-normally-incident beam of 2.275 MeV alpha particles. The normal detector angle collected particles scattered by 160° and the grazing detector was set at 104°. The atomic concentration uncertainty is ±1%. Additional details regarding the RBS measurements and analysis are given in the Supporting Information.

## Results

### Morphological and microstructural assessment of Ba-Zr-S-O thin films

First, we study the surface morphology and film microstructure of the sulfurized Ba-Zr-O thin films, summarized in **Figure 2**. We report in **Figure 2a-d** SEM images of the thin film surface at increasing sulfurization temperatures. A strong dependency of the surface morphology on the sulfurization temperature is observed, as previously reported in Ref. [17]. The thin film sample sulfurized at 700 °C (**Figure 2a**) shows a rather smooth surface when compared to films annealed at higher temperatures. On the other hand, sulfurization at 800 °C (**Figure 2b**) leads to the appearance of crystalline grains on the film surface. Further increasing the sulfurization temperature to 900 °C (**Figure 2c**) promotes the growth of more crystalline domains, characterized by dimensions smaller than 250 nm. Finally, sulfurization at 1000 °C (**Figure 2d**) leads to the formation of larger grains with dimensions up to 500 nm. To understand the origin and composition of these crystalline grains we turn to investigate the films' structural properties using TEM.

Bright-field TEM (BF-TEM) images of the cross-sections of the thin film samples are reported in **Figure 2e-h**, while **Figure 2i-l** show the selected area electron diffraction (SAED) patterns respectively acquired from the film cross-sections in **(e-h)**. **Figure S1** in the Supporting Information labels the regions from which the SAED patterns were measured.

The BF-TEM image of the film sulfurized at 700 °C (**Figure 2e**) shows that the film surface has uniform contrast, while large grains are present in the bulk of the film, as indicated by red arrows. The associated SAED pattern in **Figure 2i** shows highly discontinuous diffraction rings with multiple bright diffraction spots indicating the presence of multiple randomly oriented crystal grains of $BaZrO_3$. No clear diffraction patterns from other phases are observed at this temperature, indicating that any sulphide phase present in the film (discussed later in **Figure 3**) is likely amorphous, and thus the grains observed in **Figure 2e** are $BaZrO_3$. Based on the BF-TEM image (**Figure 2f**), sulfurization at 800 °C leads to the disappearance of the large $BaZrO_3$ grains from most of the film bulk, though $BaZrO_3$ grains remain present at the interface with the quartz glass substrate. On the other hand, strongly diffracting regions characterised by dimensions <10 nm appear in the bulk, as indicated by green arrows. The relative SAED pattern in **Figure 2j** shows the appearance of new discontinuous diffractions rings, belonging to crystalline $BaZrS_3$ and $ZrO_2$, likely linked with the appearance of nano-grains in the bulk of the film. Few bright diffraction spots belonging to the $BaZrO_3$ phase suggest that the large grains at the substrate-film interface are what remains of the original $BaZrO_3$ phase. At 900 °C, in TEM **Figure 2g** we see a complete disappearance of large $BaZrO_3$ grains at the substrate-film interface, and the appearance of considerably larger diffracting grains in the bulk of the film, as indicated by green arrows. As a consequence, the relative SAED pattern in **Figure 2k** shows brighter diffraction spots belonging to the $ZrO_2$ and $BaZrS_3$ phases. In addition, a new diffraction ring belonging to $BaZrO_3$ appears, as shown in **Figure 2k**, which most likely results from the large $BaZrO_3$ grains decomposing into smaller crystallites at the substrate-film contact.

Finally, the TEM cross-section of the sample sulfurized at 1000 °C (**Figure 2h**) shows the presence of large regions without any diffraction contrast at the surface of the film. On the other hand, strongly diffracting smaller regions with an average size of about 20 nm appear towards the interface with the substrate. In the relative SAED pattern in **Figure 2l** the presence of many diffraction spots of crystals in arbitrary orientations can be observed. Intense diffraction spots were identified to belong to $BaZrS_3$ and $ZrO_2$ crystals. The absence of rings in this SAED pattern is owed to the low number of crystal grains due to extensive crystal growth enabled by the high processing temperature.

To relate the surface morphology observed by SEM in **Figure 2(a-d)** to the film microstructure observed in TEM in **Figure 2(e-h)**, and to determine the distribution of the crystal phases observed in the SAED patterns in **Figure 2(i-l)**, we discuss next the compositional analysis of the films performed by STEM-EDX.

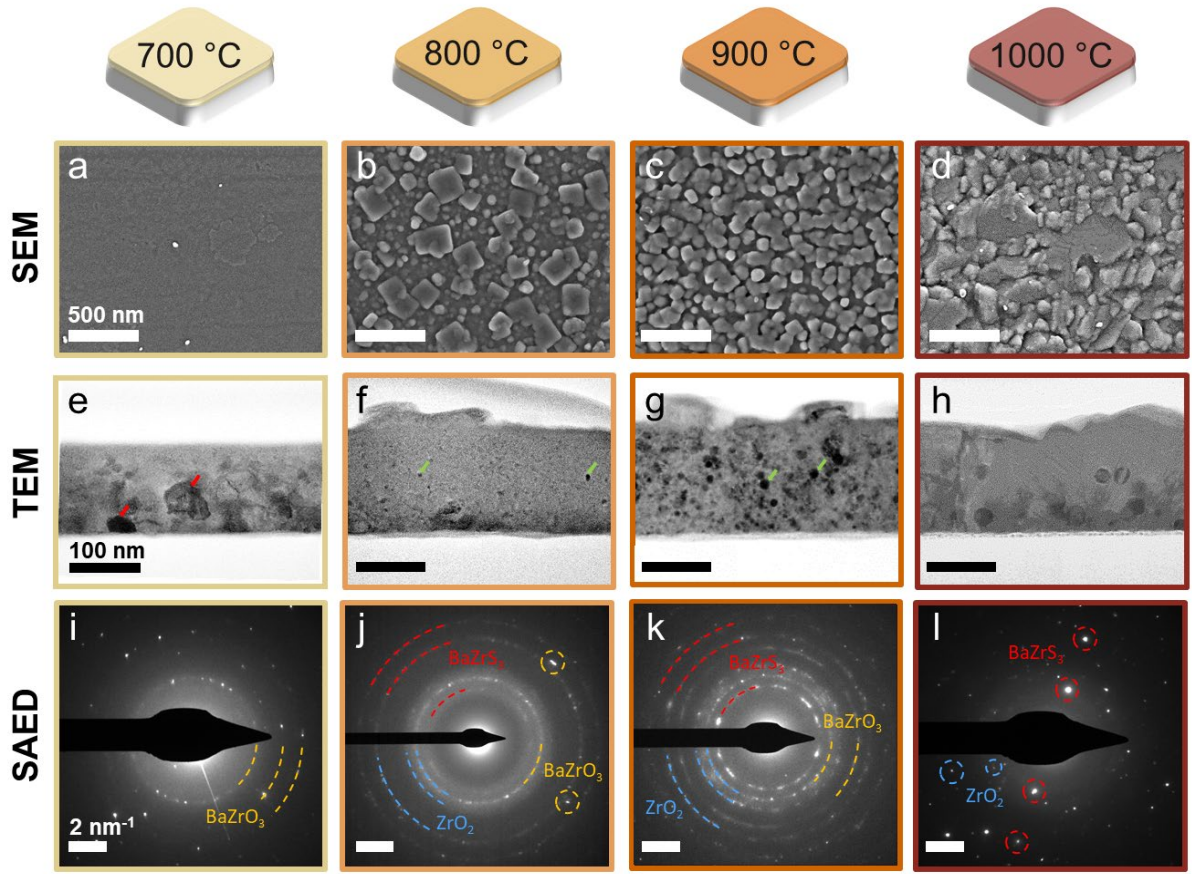

**Figure 2. Morphological and structural assessment of Ba-Zr-S-O thin films sulfurized at various temperatures:** (a-d) SEM top view images of the sulfurized Ba-Zr-O films showing an increase in surface grain density with the increase in sulfurization temperature. (e-h) Cross-sectional TEM bright field images of the thin films show changes in the microstructure. (i-l) TEM-SAED patterns obtained on the cross-sections of Ba-Zr-S-O films (avoiding diffraction from the quartz substrate) pointing to structural and phase changes with sulfurization temperature. Color labels in the SAED patterns indicate different phases: $BaZrO_3$ (orange), $BaZrS_3$ (red), $ZrO_2$ (blue). Full indexing of the diffraction planes, identification of phases, and the exact selected area used for measurements of the SAED patterns are reported in **Figure S1** in the Supporting Information.

## Compositional assessment of Ba-Zr-S-O thin films

STEM-EDX elemental mapping was performed on the cross-sections of the sulfurized thin films to obtain information on their chemical composition, as summarized in **Figure 3**. **Figure 3a** presents the ADF-STEM images and compositional mapping of the thin film cross-sections for different sulfurization temperatures. The compositional maps are superpositions of elemental compositions of cations (Ba and Zr, middle panel in **Figure 3(a)**) and anions (S and O, bottom panel in **Figure 3(a)**). Individual elemental maps of each sample are presented in the Supporting Information (**Figures S2-S5**). **Figures 3b** and **3c** show changes respectively in the cation ([Zr] / ([Zr] + [Ba])) and anion ([S] / ([S] + [O])) atomic composition as function of depth in the film, obtained from compositional line scans performed from the surface of the film towards the substrate.

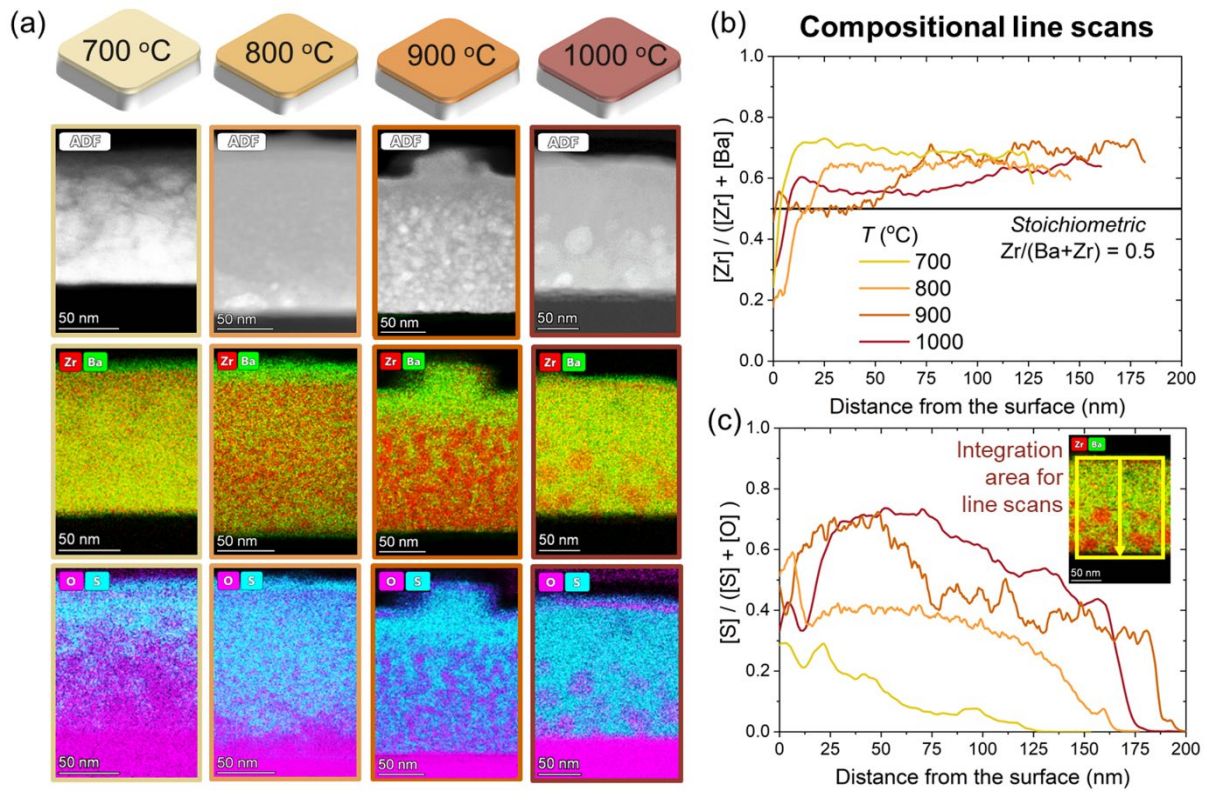

**Figure 3. Compositional assessment of Ba-Zr-S-O thin films sulfurized at various temperatures**: (a) Annular dark-field STEM images and accompanying EDX elemental maps of thin film cross-sections showing variations in Zr-Ba and O-S compositions with sulfurization temperatures. The separation between Ba-rich and Zr-rich regions in correlation with S-rich and O-rich regions is observed at higher temperatures, pointing to the formation of $BaZrS_3$ and $ZrO_2$ phases. Line scans of (b) [Zr]/([Zr] + [Ba]) and (c) [S]/([S] + [O]) show variations in cation and anion atomic compositions along the films thickness.

We first consider the average composition of the film as a function of depth. We start by highlighting compositional line scans averaged over the film width in **Figures 3b-c**. Concerning the bulk of the films, **Figure 3b** shows that all sulfurized samples are Zr-rich and Ba-poor (([Zr] / ([Zr] + [Ba]) > 0.5) due to off-stoichiometric precursor amorphous Ba-Zr-O films. [17] Regarding the anion composition, **Figure 3c** shows that, as a result of S diffusion during annealing, the S concentration is highest near the surface, and decreases towards the film-substrate interface. As expected from the increase in  diffusivity with temperature, the total S concentration in the films increases as the sulfurization temperature is increased.

However, it is interesting to note that the first 25 nm from the surface of the films shows a slightly different trend. Based on the EDX elemental maps and compositional line scans presented in **Figure 3**, we can observe Ba and S-rich layer in the first few nanometers from the surface in all thin films. Below this, there is an O-rich layer with similar thickness. This most probably indicates a formation of Ba-S-O secondary phases in the top surface layers of the films. Further characterization with XRD and Raman spectroscopy could not reveal the exact nature of the phases present in this layer, which is probably due to the very low quantity present in the film.

Next, we discuss the spatial distribution of cations and anions across the width and thickness of the films as shown in **Figure 3a**. In the film sulfurized at 700 °C, while the bulk cation composition looks fairly homogeneous, phase separation in anion composition is observed: spatial inhomogeneity indicates that separate S-rich and O-rich phases are present. This confirms the conclusions obtained from TEM analysis of this sample, which suggested the presence of an amorphous phase of Ba-Zr-S, and large crystal grains of BaZrO$_3$. On the other hand, the film sulfurized at 800 °C shows nano-phase separation of both cations and anions in correspondence with nanocrystallites seen previously in the TEM image in **Figure 2f**, and now also visible in the ADF image. These nanocrystallites correspond to a Zr- and O-rich phase, an indication of the crystalline ZrO$_2$ phase observed by TEM in SAED **Figure 2j**. Given the dim SAED spots belonging to BaZrS$_3$ in **Figure 2j** and the compositional STEM EDX contrast seen in **Figure 3a**, we believe the ZrO$_2$ crystallites are immersed in a matrix of nanocrystalline BaZrS$_3$ with possibly an amorphous component of Ba-Zr-S(-O). At the substrate-film interface, the film sulfurized at 800 °C is still strongly rich in O, confirming that the large grains observed also in **Figure 2f** are likely unreacted crystals of BaZrO$_3$. Increasing the sulfurization temperature to 900 °C leads to complete decomposition of BaZrO$_3$ grains, and to increased phase separation into O- and S-rich phases in the bulk of the film. This is in agreement with crystal growth observed from brighter spots in the SAED pattern of **Figure 2k**. Looking at the film surface, the protrusion and first 50 nm of the film show a homogeneous distribution of Ba, Zr, and S, pointing at the formation of larger grains of BaZrS$_3$. This indicates that the surface protrusions observed by SEM in **Figure 2c** are the result of the formation of BaZrS$_3$ crystals growing out of the surface. Further increasing the sulfurization temperature to 1000 °C, the STEM-EDX contrast in **Figure 3a** indicates that the sub-surface region is uniformly constituted of BaZrS$_3$, suggesting that the 500-nm-sized grains observed by SEM in **Figure 2d** are BaZrS$_3$ grains, in agreement with the bright TEM SAED spots of BaZrS$_3$ in **Figure 2l**. The size of the Zr- and O-rich phase grains also increases at this sulfurization temperature, as expected from the absence of diffraction rings in the SAED pattern of **Figure 2l**.

## Phase identification in Ba-Zr-S-O thin films

To confirm the presence of phases identified with TEM analysis in the sulfurized Ba-Zr-O thin films, we performed GIWAXS (**Figure 4**) and Raman spectroscopy (**Figure 5**).

**Figure 4a** presents representative GIWAXS patterns of sulfurized BaZrO$_3$ samples measured with grazing incidence (GI) angles ($\alpha_i$) of 0.3°, 0.5° and 2.0°, while complementary measurements with 1.0° and 1.5° are shown in **Figure S6a** in the Supporting Information. The bottom panel in **Figure 4a** shows the reference patterns of BaZrO$_3$,[30] BaZrS$_3$[30] and ZrO$_2$[31] phases, whose crystal structures are illustrated in **Figure 4b**. In the GIWAXS configuration, X-ray penetration depth is controlled by the GI angle and can be calculated based on the incident angle and material parameters as shown in Ref. [[32]]. Based on these calculations we have reported in **Figure 4c and Figure S6c** the average penetration depths of X-rays for various GI angles in the case of sulfurized BaZrO$_3$ samples. Low GI angles (0.3° and 0.5°) have an estimated penetration depth of less than 50 nm, while the higher GI angles (2°) penetrate the entire bulk of the film (> 200 nm). This means that the information obtained from GIWAXS measurements with lower GI angles is indicative mostly of the phases present near the surface of the films, while in our case at $\alpha_i = 2°$ we probe the entirety of the films.

The Bragg reflections observed in the GIWAXS patterns for sample sulfurized at 700 °C match well the reference positions of cubic perovskite-type BaZrO$_3$ (space group $Pm\overline{3}m$) for measurements with both low and high GI angles. This indicates that the crystalline BaZrO$_3$ phase is a major phase present throughout the film, as observed by TEM. Furthermore, a low intensity peak at around $2\theta = 25.3°$ is observed in the GIWAXS pattern measured with 0.3° and 0.5° GI angles. This reflection corresponds to the orthorhombic perovskite-type BaZrS$_3$ (space group $Pnma$), and indicates the formation of a small amount of crystalline BaZrS$_3$ in the surface region of the films, undetected by TEM SAED in **Figure 2i**.

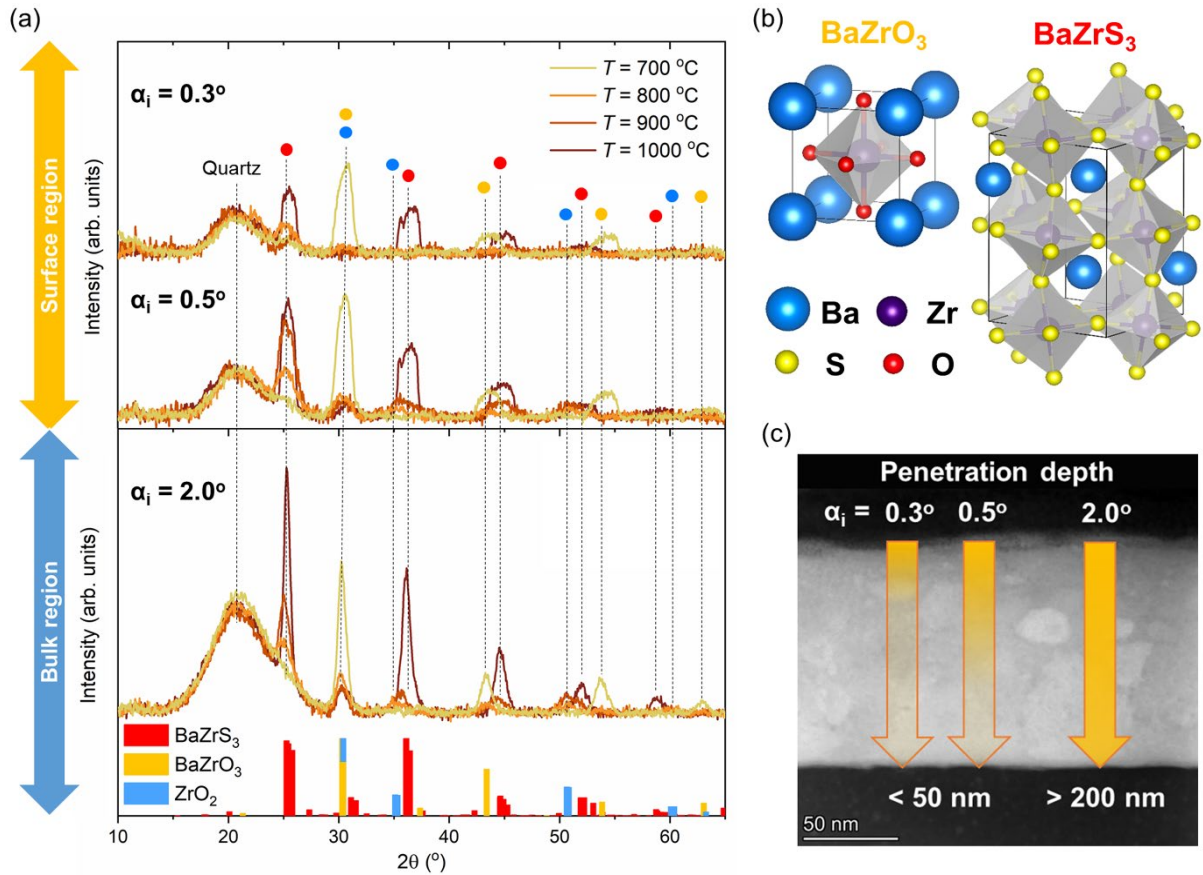

**Figure 4. Phase identification of Ba-Zr-S-O thin films sulfurized at various temperatures:** (a) Representative GIWAXS patterns of Ba-Zr-S-O films synthesized at different temperatures. The measurements were performed with grazing incidence angles ($\alpha_i$) of 0.3°, 0.5° and 2°, allowing microstructural assessment of the surface ($\alpha_i \leq 0.5°$) and bulk ($\alpha_i = 2.0°$) regions of the thin films. Based on the positions of reflections in the patterns, three phases were identified: BaZrS$_3$[30], BaZrO$_3$[30] and ZrO$_2$[31]. Reference patterns of these phases are reported on the x-axis of the GIWAXS scans. (b) Crystal structure representations of BaZrS$_3$ and BaZrO$_3$. (c) Estimated penetration depth of X-rays for different grazing incidence angles used in this study.

The samples sulfurized at 800 °C and 900 °C show less intense reflections belonging to the cubic perovskite-type BaZrO$_3$ phase, and progressively more pronounced reflections attributed to the orthorhombic perovskite-type BaZrS$_3$. Additionally, new peaks are observed at 2θ angles of 35.3° and 50.6°, which point to the formation of the ZrO$_2$ phase. The reflections belonging to the ZrO$_2$ phase could be identified with either a cubic (space group $Fm\bar{3}m$) or a tetragonal structure (space group $P42/nmc$). Differentiation between the two from the XRD measurements is difficult as XRD patterns from both structures are very similar.[33] Usually, the tetragonal structure can be distinguished by the characteristic splitting of diffraction peaks, absent in the patterns of the cubic phase. However, this task was impossible in our mixed system, with Bragg peak broadening due to small GI angles and nano-size of ZrO$_2$ crystallites. Therefore, in our case we will refer to this phase as simply ZrO$_2$.

Interestingly, we can observe that both at 800 °C and 900 °C the ratio of BaZrS$_3$ peak at 2θ = 25.3° to the oxide peak (BaZrO$_3$ or ZrO$_2$) at 2θ = 30.2° decreases towards larger GI angles, indicating that the BaZrS$_3$ phase is more prominent near the surface, as expected from the STEM analysis. Lastly, the Bragg peak corresponding to BaZrO$_3$ (2θ = 43.4°) seems to disappear in the film sulfurized at 900 °C, despite the diffused diffraction ring observed in SAED in **Figure 2k**. This is likely due to the small size of the BaZrO$_3$ crystallites, which yield a broad, non-intense peak, probably screened by the neighboring intense BaZrS$_3$ peak at $2\theta = 45.0°$. On the other hand, the Bragg peak of ZrO$_2$ ($2\theta = 50.6°$) slightly increases at 900 °C compared to 800 °C, suggesting the ZrO$_2$ phase grows at the expense of BaZrO$_3$.

As the sulfurization temperature further increases to 1000 °C, the Bragg peaks corresponding to the BaZrS$_3$ phase become more intense for all GI angles suggesting that the fraction of crystalline BaZrS$_3$ in the film increases at 1000 °C. Furthermore, at GI angle of 2°, Bragg peaks of orthorhombic perovskite-type BaZrS$_3$ are considerably sharper compared to other sulfurization temperatures, indicating more extensive BaZrS$_3$ grain growth at 1000 °C in accordance with the TEM image of **Figure 2h**. Concerning the oxide phase peaks, low-intensity peaks corresponding to the ZrO$_2$ phase are observed (mostly at $2\theta = 30.2°$), with slightly higher intensity at high GI angles of 2°. This indicates that the ZrO$_2$ phase is located mostly towards the substrate-film interface, in agreement with the STEM image in **Figure 3a**. No GIWAXS peak corresponding to BaZrO$_3$ is present at the sulfurization temperature of 1000 °C.

Raman measurements were performed complementary to the GIWAXS and TEM characterization to confirm the presence of the observed crystal and amorphous phases, the latter being especially difficult to identify with the previously used techniques.

Raman spectra were measured at 12 K using 488 nm (2.54 eV) and 532 nm (2.33 eV) laser excitations. Low-temperature conditions were chosen due to the increase in the phonon lifetime, which reduces the widths of the peaks, yielding a better resolution. Two excitation lasers were used to differentiate the probed volume of the sample. Due to the stronger absorption at short wavelengths the Raman signal recorded with the 488 nm laser is weighted more toward the surface region (estimated penetration depth < 100 nm), than the Raman signal recorded with the 532 nm laser with an estimated penetration depth < 150 nm for our 200nm thick sample.

The bottom panel in **Figure 5** presents Raman peak positions of the BaZrS$_3$,[34,35] BaZrO$_3$,[36,37] and ZrO$_2$[33,38] structures from literature, which are used for the identification of phases in the synthesized layers.

It should be noted that BaZrO$_3$ with a perfect cubic perovskite structure (space group $Pm\bar{3}m$) does not have any active Raman modes.[39] Therefore, it has been debated if the presence of peaks in the Raman spectrum of BaZrO$_3$ actually signals the lowering of symmetry due to local distortions, indicating that the material is cubic only on average or if these peaks actually correspond to second-order Raman scattering.[40–44] Recent results from neutron scattering,[45] as well as detailed experimental and theoretical Raman study of BaZrO$_3$ have shown that there is no lowering of cubic symmetry.[37,46] This means that the peaks observed in the Raman spectrum of BaZrO$_3$ are indeed of second order and not related to structural distortions. In difference to the first order Raman scattering, second-order involves two phonons with opposite parallel wave-vectors in the same scattering event. This means that the conservation of momentum for second-order processes is always satisfied, and second-order Raman scattering is always allowed. Considering the involvement of two phonons, second-order Raman scattering is less probable than the first-order, and therefore usually weaker in intensity. However, in the absence of the first-order scattering, it can become dominant in the Raman spectra, as is the case with BaZrO$_3$.

The Raman spectra measured with 532 and 488 nm excitation, of the sample sulfurized at 700 °C is mostly dominated by broad peaks belonging to the BaZrO$_3$ phase. Furthermore, lower intensity peaks corresponding to the BaZrS$_3$ phase are observed. The very broad characteristics of the BaZrS$_3$ peaks at 700 °C suggest the phase is (mostly) amorphous, as hypothesized from the TEM SAED patterns of the sample in **Figure 2i**.

In agreement with the GIWAXS measurements, at the sulfurization temperature of 800 °C we observe a higher intensity of the peaks corresponding to the broad amorphous/nanocrystalline BaZrS$_3$ phase, and the appearance of a small peak at 320 cm$^{-1}$ corresponding to ZrO$_2$. An increase in the sulfurization temperature to 900 °C leads to an increase in the intensity of Raman peaks corresponding to BaZrS$_3$, and a reduction in the intensity of peaks belonging to $BaZrO_3$, corresponding to an increase in the fraction of $BaZrS_3$ phase over $BaZrO_3$. Furthermore, additional Raman peaks belonging to $ZrO_2$ phase are observed in the spectra measured under both excitations, implying the formation of this phase throughout the thickness of the layer for a sulfurization temperature of 900 °C. Finally, sulfurization at 1000 °C leads to the appearance of more intense $BaZrS_3$ Raman peaks compared to other sulfurization temperatures. It should be noted that the widths of the Raman peaks belonging to the one-phonon processes of $BaZrS_3$ phase are comparable to the instrumental broadening of the Raman system of ≈ __ cm-1. This indicates that the formed $BaZrS_3$ phases are highly crystalline. As expected from GIWAXS and TEM characterization, Raman peaks belonging to $ZrO_2$ phase are still present in the spectra while no significant contributions from $BaZrO_3$ are observed. Furthermore, Raman spectra of sample sulfurized at 1000 °C do not show any clear evidence of the presence of leftover broad amorphous peaks.

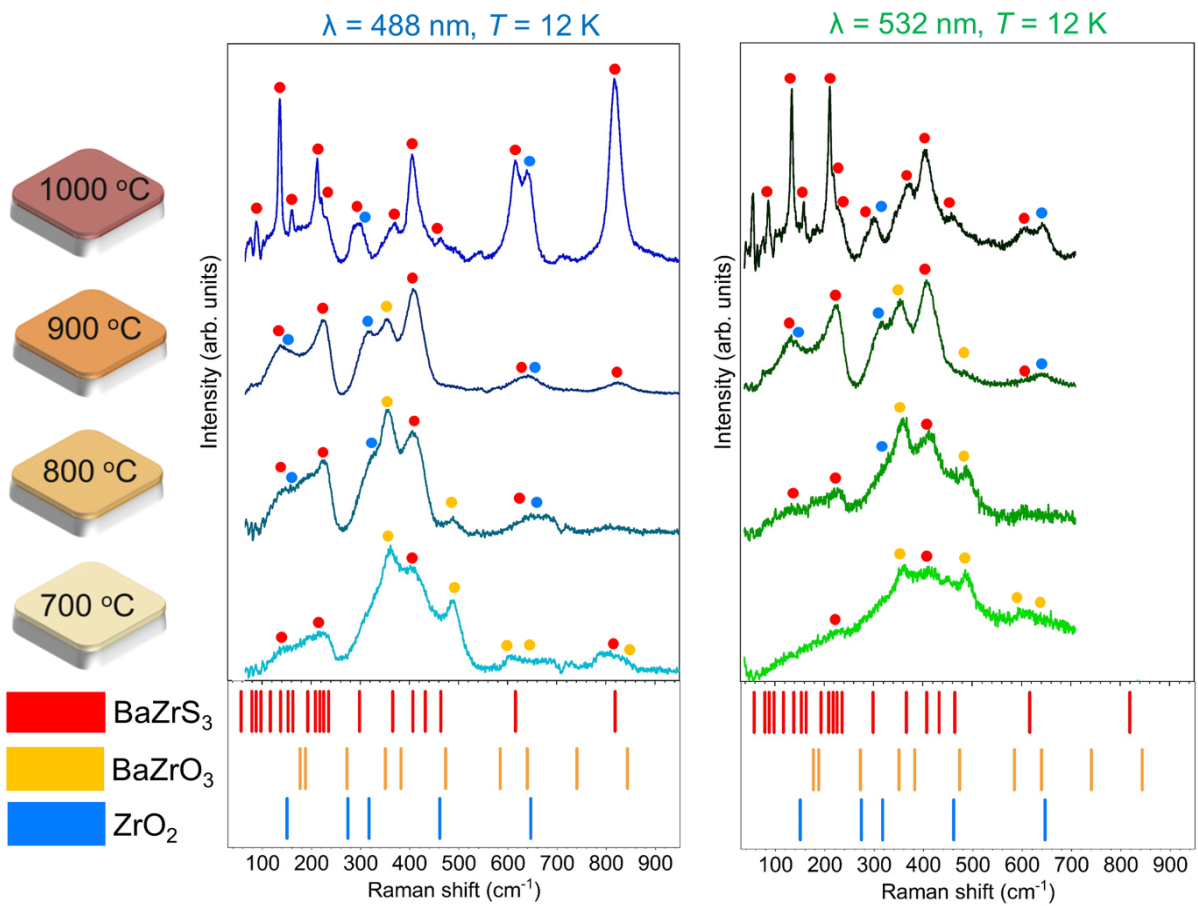

**Figure 5. Raman characterization of Ba-Zr-S-O thin films sulfurized at various temperatures:** Raman spectra of the four Ba-Zr-S-O thin films measured at 12 K with (left) 488 and (right) 532 nm excitations. Identification of Raman peaks is performed based on the reference Raman peak positions belonging to (red) $BaZrS_3$,[34,35] (orange) $BaZrO_3$,[36,37] and (blue) $ZrO_2$[33,38] phases, shown under the measured Raman spectra.

## Compositional stoichiometry determination of the $BaZrS_3$ phase

The Ba-Zr-S-O thin film sulfurized at 1000 °C is used as a representative sample to probe the compositional stoichiometry of the $BaZrS_3$ phase. The RBS spectrum of this sample is shown in **Figure S7** in the **Supporting Information** along with the fits corresponding to different element contributions, including Ba, Zr, S, O and Si. Based on this, we were able to determine the elemental composition of the Ba-Zr-S-O thin film after subtracting the $SiO_2$ composition, corresponding to the quartz substrate. The results are shown in the second column in **Table 1**. These results correspond to the integral composition of the overall thin film, and are also in good agreement with the EDX measurements. It should be noted that the obtained results in this study for the anion composition match well the previous compositional analysis reported in Ref. [[17]], however there is a slight discrepancy in the measured cation composition. Considering that the EDX and RBS measurements in this study match very well, we tend to believe that the values here represent more accurate assessment of the films.

**Table 1.** Compositional profiling of the Ba-Zr-S-O thin film sulfurized at 1000 °C obtained from the RBS measurements. The values are used to determine the stoichiometry of the formed $BaZrS_3$ phase.

| Element | Composition of the Ba-Zr-S-O thin film (%at) | Fraction of the composition corresponding to $ZrO_2$ (%at) | Fraction of the composition corresponding to $BaZrS_3$ (%at) |
|---|---|---|---|
| **Ba** | 15.7±1.0 | 0 | 15.7±1.0 |
| **Zr** | 22.8±1.0 | 7.1±1.0 | 15.7±1.0 |
| **S** | 47.2±1.0 | 0 | 47.2±1.0 |
| **O** | 14.3±1.0 | 14.3±1.0 | 0 |

Considering that SAED, GIWAXS and Raman measurements point to the presence of only $BaZrS_3$ and $ZrO_2$ crystalline phases within the sample, we can determine the stoichiometry of $BaZrS_3$, by subtracting the composition of $ZrO_2$. It should be noted that $ZrO_2$ can deviate from the perfect stoichiometry of Zr:O = 1:2 towards oxygen-poor compounds $ZrO_{2-x}$, which is usually seen in the Raman spectra by the appearance of significant broadening of the peaks.[47] We do not observe such broad peaks in the measured Raman spectra, which point to the formation of $ZrO_2$ with a composition close to stoichiometry. Assuming that all of O is used for the formation of $ZrO_2$, which is corroborated by SAED, GIWAXS and Raman measurements, we can calculate how much of total Zr participates in $ZrO_2$, as shown in the third column in Table 1.

Subtracting the composition of $ZrO_2$ from the RBS results, we obtain the $BaZrS_3$ composition as 15.7±1.0 %at of Zr, 15.7±1.0 %at of Ba, 47.2±1.0 %at of S (fourth column in **Table 1**). This points to the formation of $BaZrS_3$ phase with a perfect stoichiometry of Ba:Zr:S = 1:1:3.

## Discussion

Before discussing the sulfurization reaction mechanisms, we highlight that (i) at sulfurization temperatures of 700°C an amorphous oxysulfide Ba-Zr-S-O phase seems to be present. Given the gradual increase in $BaZrS_3$ GIWAXS peak intensity, we could expect this amorphous phase to be present in gradually smaller fractions as the processing temperature is increased. (ii) Sulfurization of Ba-Zr-O at different temperatures does not lead to formation of a crystalline solid-solution $BaZr(O_xS_{1-x})_3$, confirming the results from Ref. [[17]]. This is shown by both Raman spectroscopy and GIWAXS, as the measured peaks are always centered at the expected positions of either the oxygen-pure $BaZrO_3$ phase or the sulfur-pure $BaZrS_3$ phase, and it is confirmed by RBS. This means that the synthesized layers present a mixture of crystalline domains corresponding to either cubic $BaZrO_3$, orthorhombic $BaZrS_3$, or other secondary phases, while no domains are present where S and O are statistically intermixed within a crystalline single phase. This agrees with recent density functional theory calculations on thermodynamic stability of $BaZr(O, S)_3$ phase, which predicts the solid solution $BaZr(O,S)_3$ phase to be thermodynamically unstable and easily decomposed into secondary phases such as $ZrX$ , $BaX$, $ZrX_2$ and $ZrX_2$ (X = S or O) or directly phase-separating into $BaZrO_3$ and $BaZrS_3$.[48]

We now propose the following formation mechanism of the sulfide perovskite phase from amorphous Ba-Zr-O precursors, based on the morphological and microstructural characterization of the films sulfurized at different temperatures. A schematic overview of the reaction process under different annealing conditions and its products is presented in **Figure 6**. From the results of Refs. [[49,50]], we expect that during heating the amorphous precursor Ba-Zr-O film crystallizes in the cubic perovskite-type $BaZrO_3$ crystal phase already at temperatures below 700 °C. As the film is exposed in $H_2S$ atmosphere for 30 minutes at 700 °C, $BaZrO_3$ grains grow further, while S species diffuse into the film. Judging from the S-rich and O-rich phase separation observed by STEM ADF in **Figure 3a** (top-left), we speculate that S diffuses through the $BaZrO_3$ grain boundaries, reacting at the grains surfaces. Here, the absence of crystalline STEM ADF contrast between $BaZrO_3$ grains suggests that an amorphous Ba-Zr-S-O phase forms upon the reaction of $BaZrO_3$ with $H_2S$. At 700 °C, $BaZrO_3$ grains are almost completely converted to amorphous Ba-Zr-S-O on the film surface, where S is present in considerably higher concentrations with respect to the bulk. GIWAXS and Raman measurements also point at initial $BaZrS_3$ crystal growth at the film surface. At 800 °C, S diffusion occurs more rapidly, and the total concentration of S in the film increases. As a consequence, the $BaZrO_3$ crystal phase is amorphized to a larger extent than at 700 °C, and large $BaZrO_3$ grains remain only at the substrate-film interface, as seen in **Figure 2f**. At this higher sulfurization temperature, signatures of $BaZrS_3$ crystal phase are present from the entire film in both GIWAXS and TEM SAED. In addition, at 800 °C a new nanocrystalline phase of $ZrO_2$ forms in the bulk of the film, evidenced by GIWAXS, Raman, STEM and TEM SAED. Because the initial amorphous Ba-Zr-O precursor film was rich in Zr, as shown in the compositional analysis of **Figure 3b**, the formation of the $ZrO_2$ may be attributed to capturing of out-diffusing O from the extra, off-stoichiometric Zr present in the film. Zr would capture preferentially O rather than S because $ZrS_2$ has higher formation energy (-1.956 eV/atom)[51] than $ZrO_2$ (-3.755 eV/atom).[52] At 900 °C, the $BaZrO_3$ phase almost vanishes, being present only in the form of nanocrystallites observed in the TEM SAED pattern of **Figure 2k**. At this temperature, XRD and TEM SAED data suggest an increased grain growth of both the $BaZrS_3$ and $ZrO_2$ phases. Finally, at 1000 °C the $BaZrO_3$ phase is absent, while crystal grains of both $ZrO_2$ and $BaZrS_3$ grow substantially larger compared to the lower sulfurization temperatures, yielding sharper peaks in both GIWAXS and Raman spectroscopy. $BaZrS_3$ grains grow within the film bulk and out of the film surface, as seen by both SEM and TEM respectively in **Figure 2b-d** and **Figure 2f-h**. Growth out of the surface likely occurs because of the lower atomic density of the $BaZrS_3$ orthorhombic perovskite-type crystal structure compared to the cubic $BaZrO_3$. Judging from the final concentration of S in the film in **Figure 3c**, and the absence of amorphous phases at 1000 °C, we can infer most of the initial O present in the film has diffused out during sulfurization.

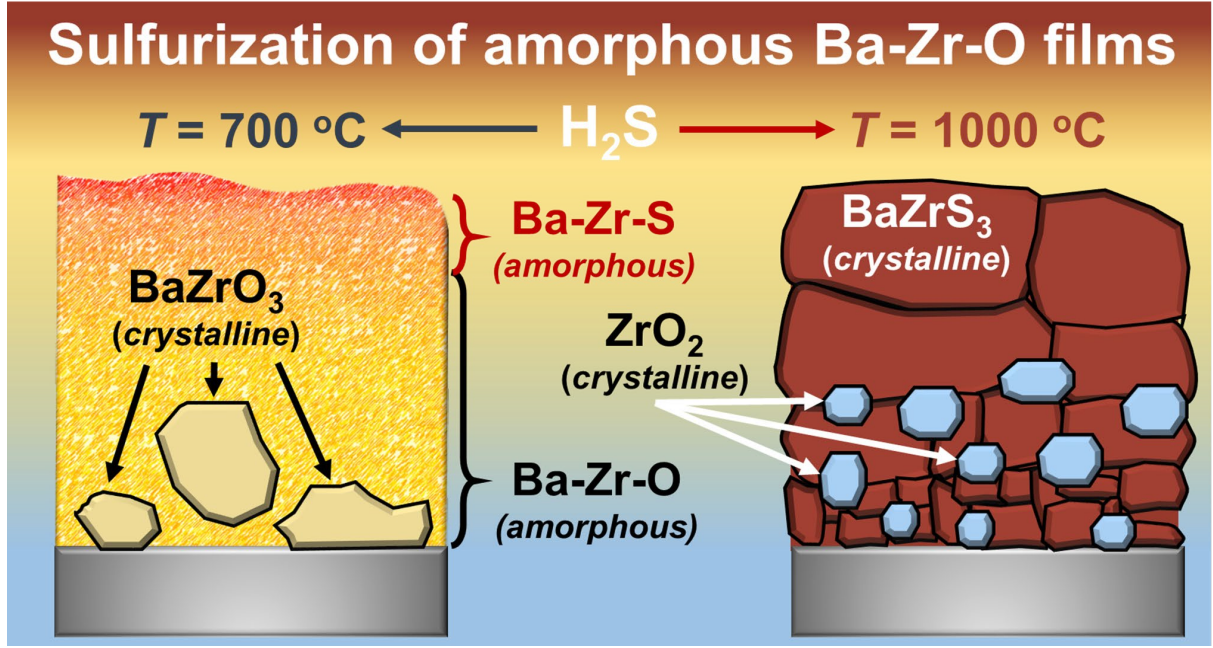

**Figure 6.** Schematic overview of the sulfurization process of amorphous Ba-Zr-O films with indications of the main phases formed at annealing temperatures of 700 and 1000 ºC.

Considering the reactions occurring during sulfurization at different temperatures, we discuss the rate-limiting factors for the formation of BaZrS₃ by sulfurization of BaZrO₃. Several factors influence, and can possibly limit this reaction: (i) the reaction energetic barrier involved in converting oxides to sulfides, (ii) the diffusion of S atoms in BaZrO₃ layers and (iii) the preferential formation of secondary phases. In perfectly stoichiometric compositional conditions, BaZrS₃ is expected to form following the reaction

$$c\text{BaZrO}_{3(s)} + 3\text{H}_2\text{S}_{(g)} \rightarrow c\text{BaZrS}_{3(s)} + 3\text{H}_2\text{O}_{(g)} \tag{1}$$

Based on our experimental results, we observe that this reaction goes in fact through an intermediate amorphization step of the BaZrO₃ crystal phase. Reaction (1) can thus be expressed as a 2-step reaction (considering the case of stoichiometric cation (Ba and Zr) compositions):

$$c\text{BaZrO}_{3(s)} + 3\text{H}_2\text{S}_{(g)} \rightarrow a\text{BaZrS}_{3(s)} + 3\text{H}_2\text{O}_{(g)} \tag{2}$$

$$a\text{BaZrS}_{3(s)} \rightarrow c\text{BaZrS}_{3(s)} \tag{3}$$

From **Figure 3a**, we can observe that already at 700 °C no large BaZrO₃ crystal grains are present at the surface, where the S concentration is high, indicating that S amorphized the BaZrO₃ crystal phase on surface (reaction (2)). On the other hand, in the bulk of the film, where only a small amount of S could diffuse, we can still observe large BaZrO₃ grains. From this, we can infer that reaction (2) can occur already at 700 °C wherever enough S is present. In addition, results from Refs [16, 18] show that reaction (3) can occur already at temperatures as low as 550 °C. The presence of Ba-Zr-S-O amorphous regions may thus be explained with the S concentration in these regions being too low for crystallization to stoichiometric BaZrS₃, as shown in the EDX elemental line scans in **Figure 3c**. These observations allow to conclude that the diffusion of S species into the film is the rate-limiting step of the transformation of $c$BaZrO₃ into $c$BaZrS₃ described by reaction (1). Increasing the processing temperature allows to increase S diffusion into the film, and thus the fraction of formed BaZrS₃ crystalline phase. Indeed, our characterization shows that the BaZrS₃ crystalline phase grows in volume as the processing temperature is increased.

We must consider that the formation of the $ZrO_2$ secondary phase may also be a limiting factor in reaction (1), as O may be capturing Zr, decreasing the yield of reaction. We suspect this is not the case in our system, as the $ZrO_2$ phase in our samples could form due to the initial Zr-rich amorphous precursor film composition as described in a hypothetical global formation process presented in reaction (4).

$$BaZr_{1+x}O_{3+2x(s)} + 3H_2S_{(g)} \rightarrow BaZrS_{3(s)} + 3H_2O_{(g)} + xZrO_{2(s)} \qquad (4)$$

However, to confirm this, sulfurization of Zr-poor precursor films should be investigated. The absence of $ZrO_2$ formation in these conditions would confirm that $ZrO_2$ forms as a result of O capture from extra, off-stoichiometric Zr. On the other hand, the presence of $ZrO_2$ in a sulfurized Zr-poor Ba-Zr-O film would indicate that the out-diffusing O is partly capturing Zr from the film to form $ZrO_2$. In the latter case, this step would be a limiting factor of reaction (1).

Finally, we comment on the formation of the stoichiometric $BaZrS_3$ phase (Ba:Zr:S = 1:1:3) observed by complementary GIWAXS, Raman and RBS characterization, despite the Zr-rich nature of the precursor amorphous Ba-Zr-O film. The absence of off-stoichiometric BaZrS agrees with density functional theory calculations on intrinsic defects in $BaZrS_3$, which showed that $BaZrS_3$ is sufficiently defect-tolerant due to the high formation energies of compositional anion and cation defects.[53] Remarkably, this is in difference to other chalcogenide systems, such as chalcopyrites (Cu(In, Ga)Se$_2$) or kesterites (Cu$_2$ZnSn(S,Se)$_4$) which tend to accommodate quite a wide range of off-stoichiometric compositions.[26,28] While on one hand the stoichiometric compositional growth of $BaZrS_3$ can be beneficial due to avoidance of possibly detrimental compositional defects, on the other hand, it can also lead to limitations in the tunability of properties that could be achieved by changes in the composition.

## Conclusions

We have investigated a series of Ba-Zr-O thin films sulfurized in an Ar + 5%H$_2$S atmosphere for 30 minutes at temperatures ranging from 700 °C to 1000 °C. We used STEM-EDX, TEM SAED, GIWAXS and Raman spectroscopy to investigate in detail the effect of sulfurization on the morphological, compositional and structural properties of the formed layers. We observed the conversion of the initial oxide precursor phase into a crystalline $BaZrS_3$ phase and a $ZrO_2$ secondary phase, the volume fraction of these phases being strongly influenced by the sulfurization temperature. We propose a formation mechanism of $BaZrS_3$ based on a two-step reaction involving (1) an intermediate amorphization step of the $BaZrO_3$ crystalline phase, with (2) subsequent crystallization into $c$BaZrS$_3$ of the intermediary amorphous Ba-Zr-S-O phase. We show that the diffusion of S into the film is the rate-limiting step of the sulfurization of the Ba-Zr-O precursor film, explaining the observed increase in $c$BaZrS$_3$ volume fraction with increasing processing temperature. At the highest sulfurization temperature used in this study (1000 °C), we obtained ~500-nm-sized $BaZrS_3$ crystalline grains. Although synthesized from a Zr-rich precursor film, the film consisted of only $BaZrS_3$ stoichiometric crystal grains and a small amount of $ZrO_2$ phase. Thus, our study reveals $BaZrS_3$ as quite unique among the other types of chalcogenides, e.g. chalcopyrite and kesterites, because it can be stoichiometrically synthesized from precursors with elemental compositions deviating from the 1:1:3 stoichiometry.

## Supporting Information

Supporting Information is available free of charge at XXXX.

STEM-EDX elemental maps, GIWAXS patterns, Raman measurements, SAED-TEM patterns

## Competing Interests

The authors declare that they have no competing financial interests.

## Authors contributions

M.D. and J.M.P. conceived the research idea and supervised the work, with the help from A.F.M. and T.U. J.M.P. and M.R. prepared the samples. S.R.P. and A.G. prepared the thin film lamellas, and preformed the STEM-EDX analysis. E.S., B.X.M.R., and I.T.F.M.L. performed Raman measurements and analysis. M.D. and S.S. did GIWAXS experiments and analysis. M.D., A.G., and S.R.P. wrote the manuscript with the inputs from all authors.

## Acknowledgments

The authors gratefully acknowledge support from the Secretariat of Education, Research and Innovation (SERI) to fund the Horizon Europe EIC PathFinder Open project SOLARUP (project number 101046297). M.D. thanks funding from H2020 through the Marie Curie Project SMARTCELL (project number 101022257).

**Supporting Information for:**

**Understanding the growth mechanism of BaZrS$_3$ chalcogenide perovskite thin films from sulfurized oxide precursors**

Santhanu Panikar Ramanandan,[1§] Andrea Giunto,[1§] Elias Z. Stutz,[1] Benoit Xavier Marie Reyner,[1] Iléane Tiphaine Françoise Marie Lefevre,[1] Marin Rusu,[2] Susan Schorr,[2,3] Thomas Unold,[2] Anna Fontcuberta i Morral,[1,4] José Márquez Prieto,[2,5]* Mirjana Dimitrievska,[1,6]*

*1 - Laboratory of Semiconductor Materials, Institute of Materials, Faculty of Engineering, Ecole Polytechnique Fédérale de Lausanne (EPFL), 1015 Lausanne, Switzerland.*

*2 - Department of Structure and Dynamics of Energy Materials, Helmholtz-Zentrum Berlin für Materialien und Energie GmbH, Hahn-Meitner-Platz 1, 14109 Berlin, Germany.*

*3 – Institute of Geological Sciences, Freie Universitaet Berlin, Maltese St. 74-100, 12249 Berlin, Germany*

*4 - Institute of Physics, Faculty of Basic Sciences, Ecole Polytechnique Fédérale de Lausanne, 1015 Lausanne, Switzerland.*

*5 - Humboldt University of Berlin, Unter den Linden 6, 10117 Berlin, Germany.*

*6 - Transport at Nanoscale Interfaces Laboratory, Swiss Federal Laboratories for Material Science and Technology (EMPA) Ueberlandstrasse 129, 8600 Duebendorf, Switzerland*

[§]These authors have contributed equally to this work

*corresponding authors:* jose.marquez@physik.hu-berlin.de; mirjana.dimitrievska@empa.ch

**Structural assessment of Ba-Zr-S-O thin films sulfurized at various temperatures: TEM-SAED analysis**

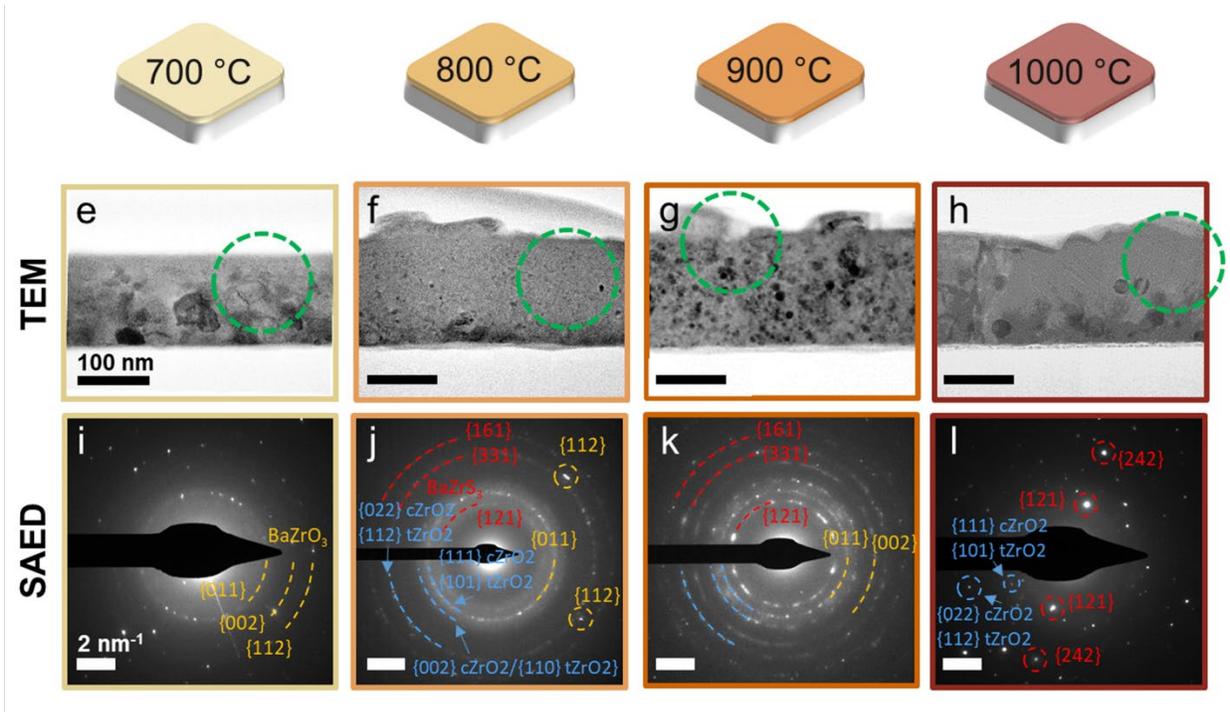

**Figure S1. Structural assessment of Ba-Zr-S-O thin films sulfurized at various temperatures:** (e-h) Cross-sectional TEM bright field images of the thin films showing changes in the microstructure. Green circles indicate areas from which SAED patterns were measured. **(i-l)** TEM-SAED patterns obtained from the labeled areas on the cross-sections of Ba-Zr-S-O films, along with the full indexing of the diffraction planes and identification of phases. Color labels in the SAED patterns indicate different phases: BaZrO₃ (orange), BaZrS₃ (red), ZrO₂ (blue). It should be noted that the exact structure of the ZrO₂ phase could not be distinguished from the SAED measurements, due to the similarity in patterns corresponding to cubic (space group *Fm*-3*m*) (labeled cZrO₂ in **Figure (i-l)**) and tetragonal structure (space group *P*42/*nmc*) (labeled tZrO₂ in **Figure (i-l)**).

**Compositional assessment of Ba-Zr-S-O thin films sulfurized at various temperatures: STEM-EDX analysis**

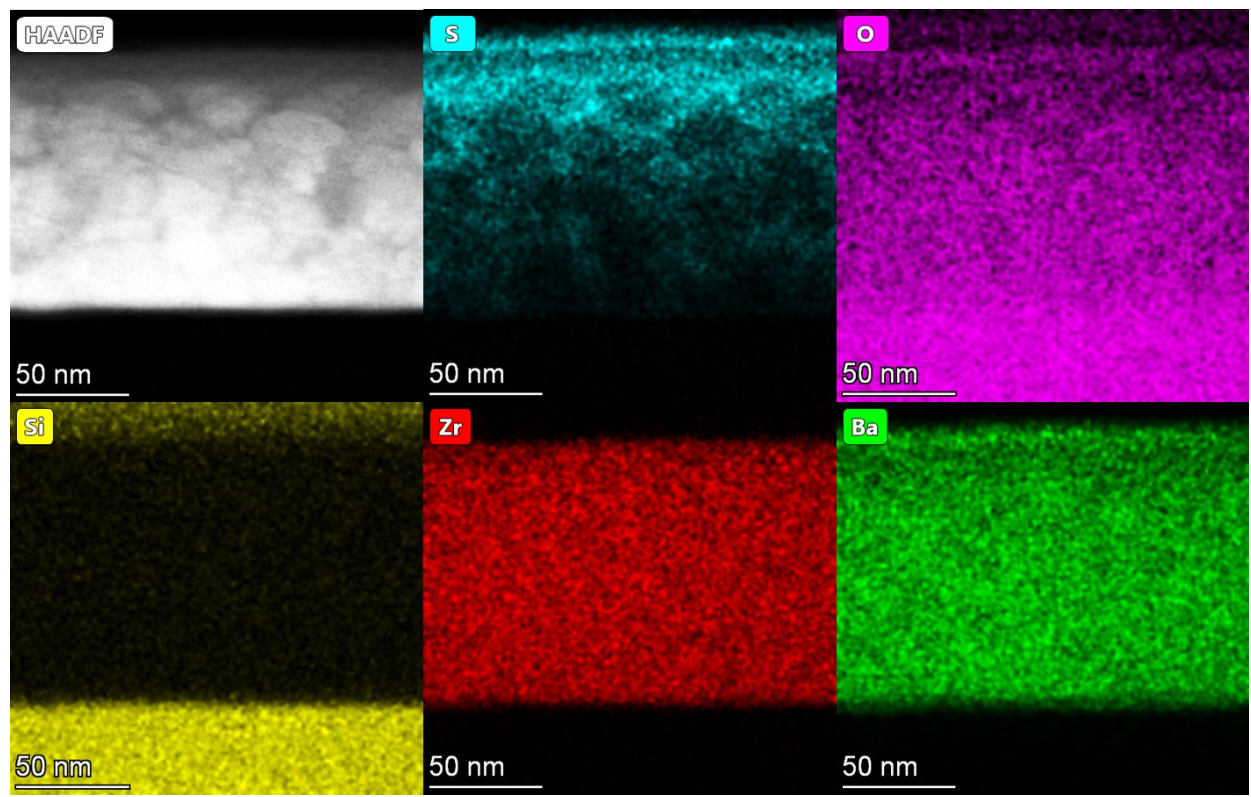

**Figure S2. Compositional assessment of Ba-Zr-S-O thin films sulfurized at 700 ºC.** Annular dark-field STEM images and accompanying EDX elemental maps of thin film cross-sections showing variations in Zr, Ba, O, S and Si compositions.

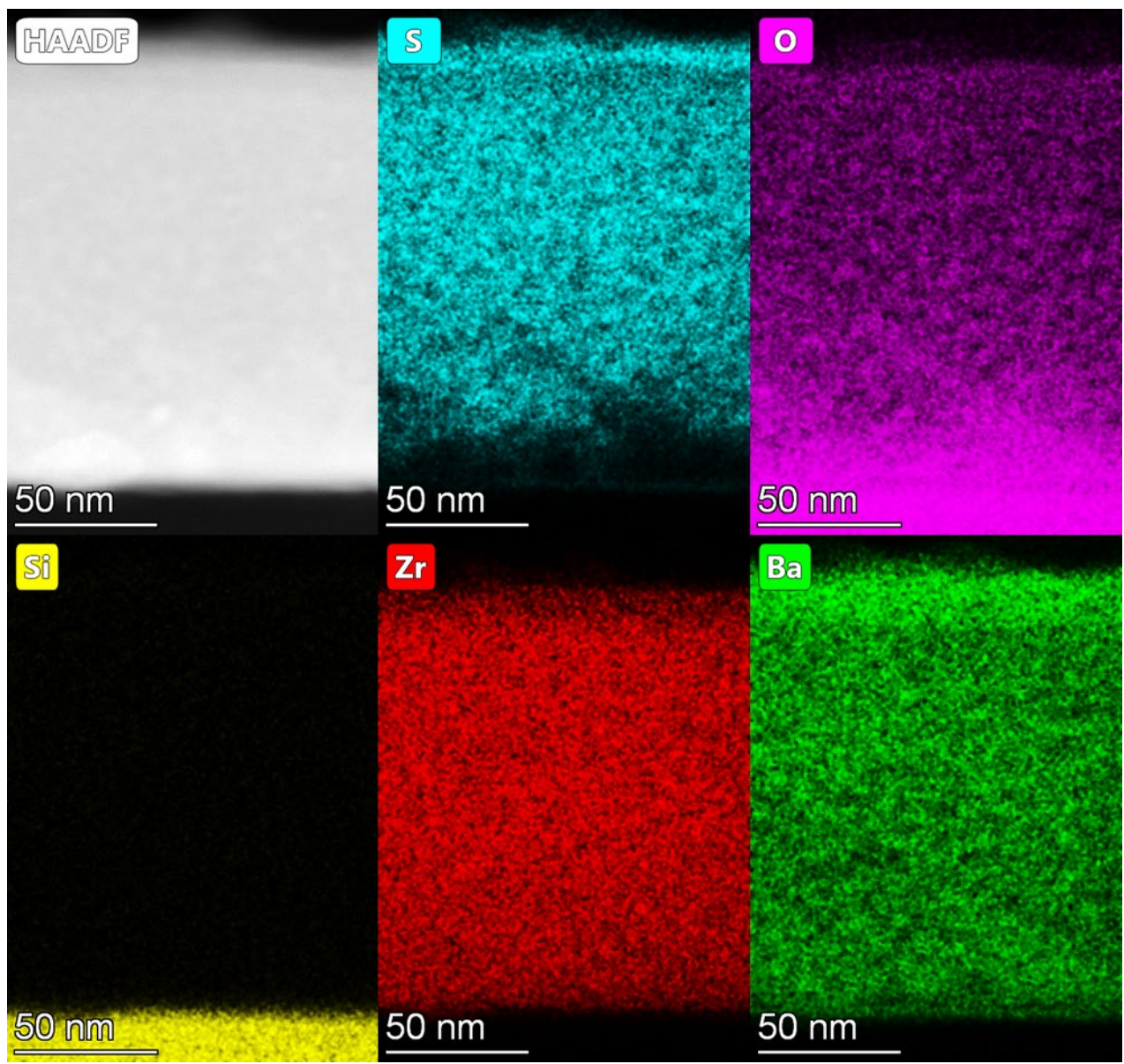

**Figure S3. Compositional assessment of Ba-Zr-S-O thin films sulfurized at 800 ºC.** Annular dark-field STEM images and accompanying EDX elemental maps of thin film cross-sections showing variations in Zr, Ba, O, S and Si compositions.

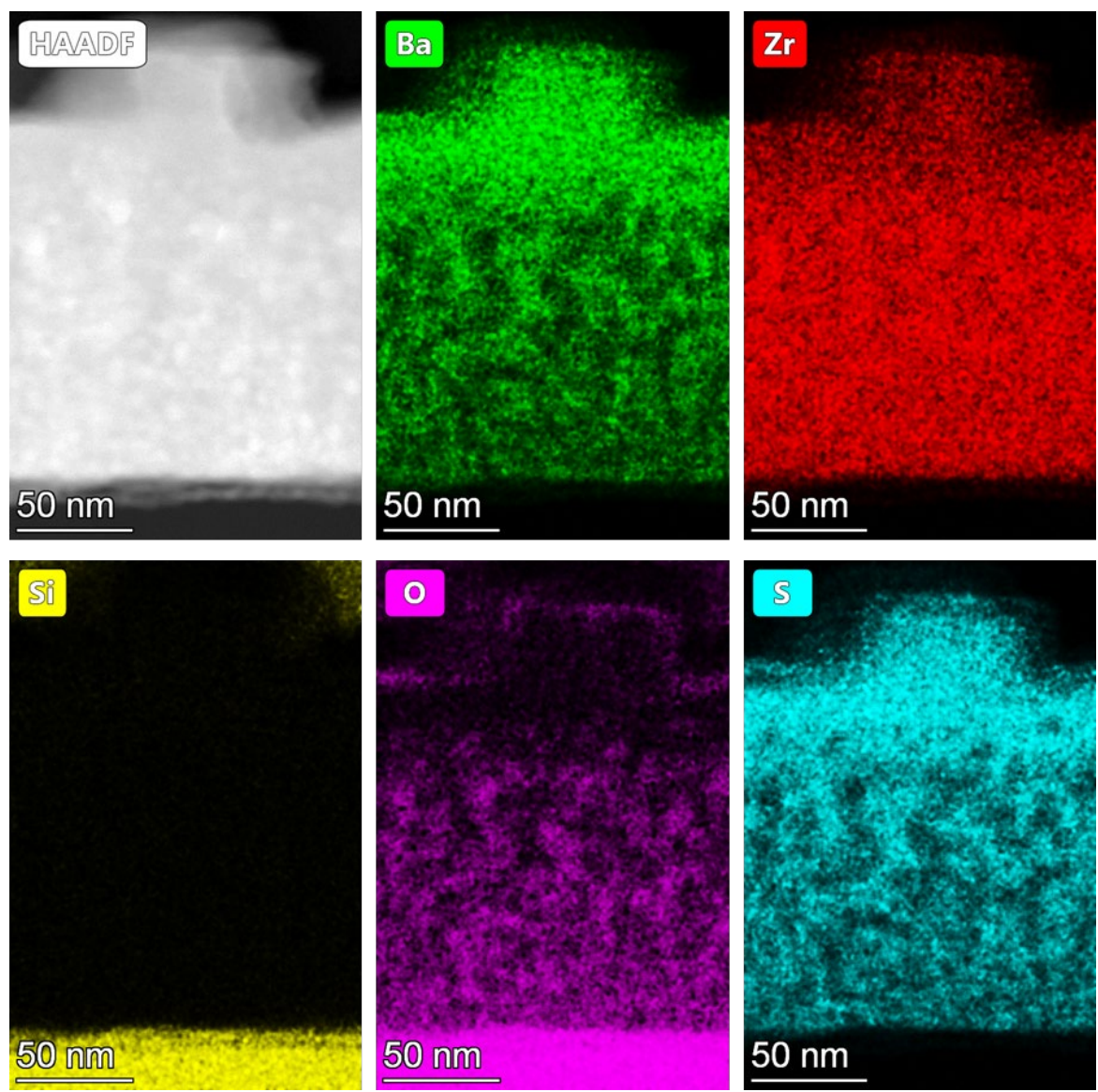

**Figure S4. Compositional assessment of Ba-Zr-S-O thin films sulfurized at 900 ºC**. Annular dark-field STEM images and accompanying EDX elemental maps of thin film cross-sections showing variations in Zr, Ba, O, S and Si compositions.

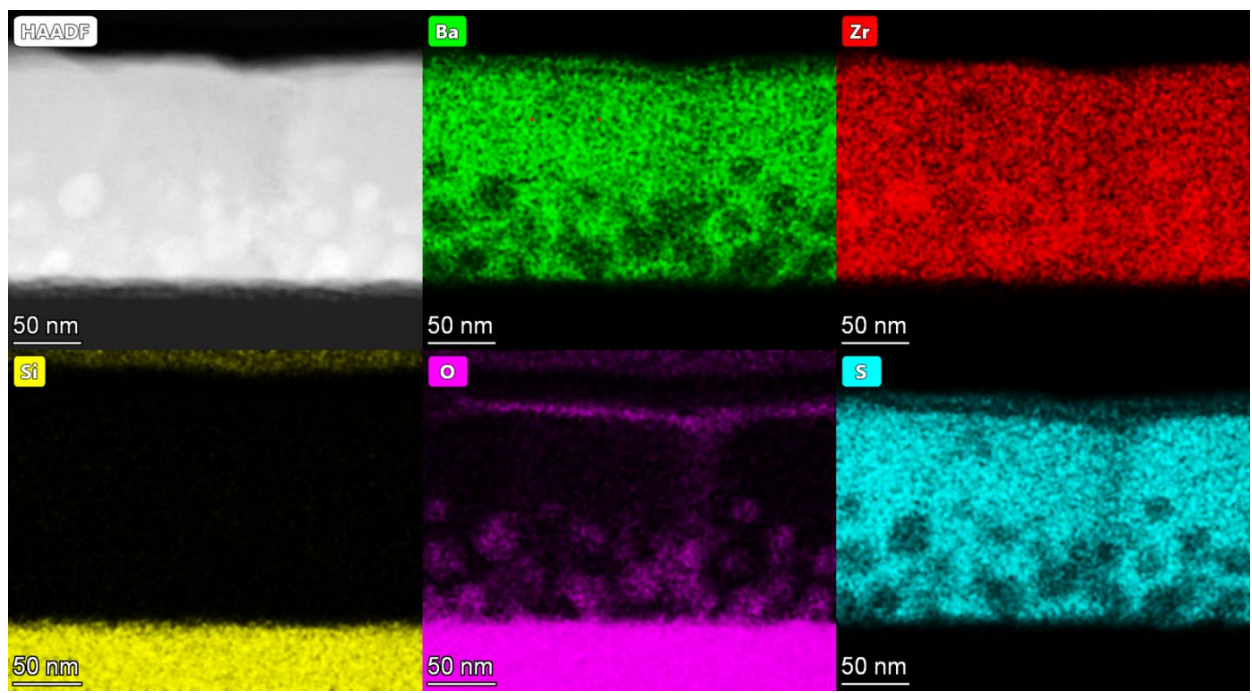

**Figure S5. Compositional assessment of Ba-Zr-S-O thin films sulfurized at 1000 ºC.** Annular dark-field STEM images and accompanying EDX elemental maps of thin film cross-sections showing variations in Zr, Ba, O, S and Si compositions.

**Phase identification in Ba-Zr-S-O thin films: additional GIWAXS measurements**

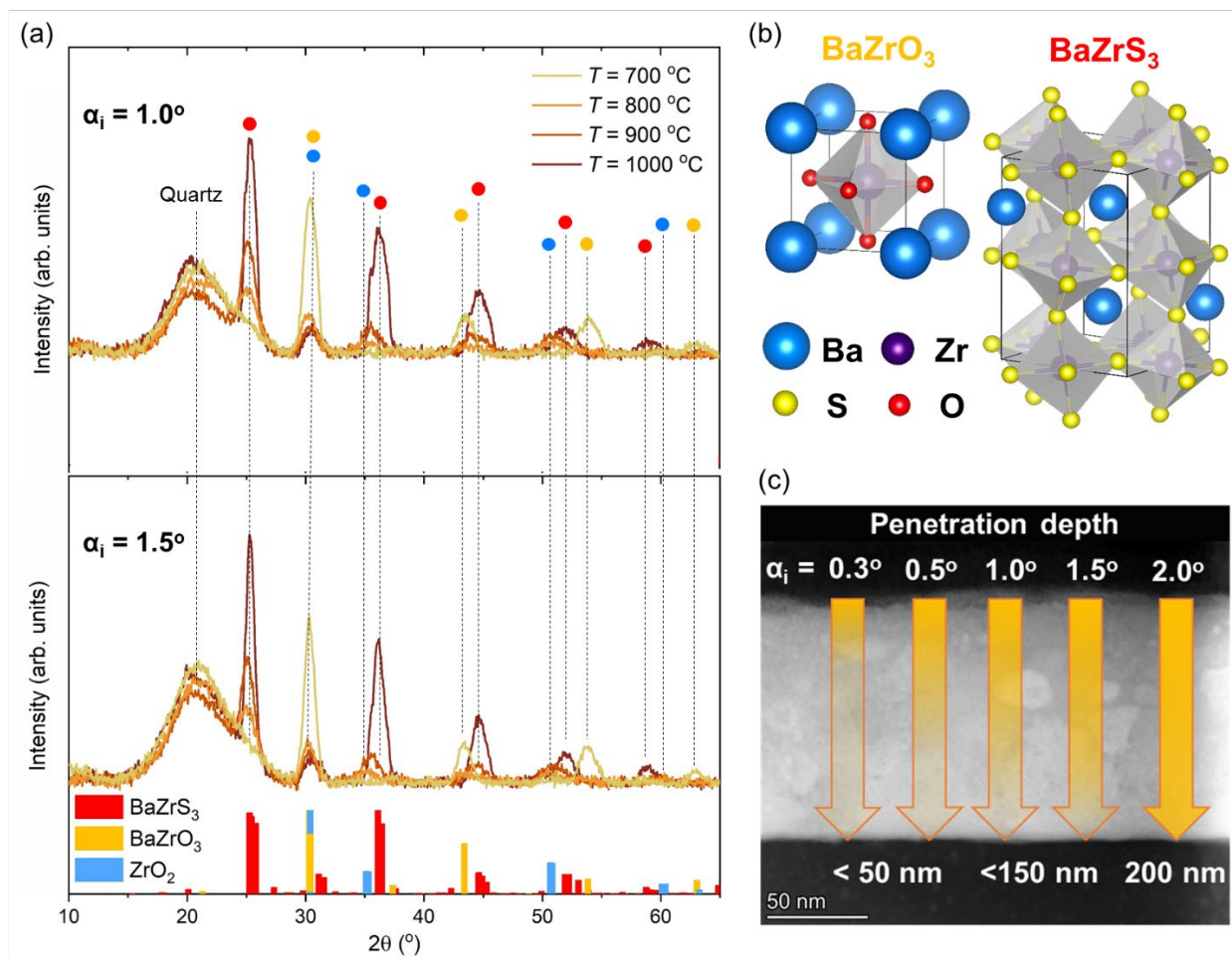

**Figure S6. Phase identification of Ba-Zr-S-O thin films sulfurized at various temperatures:** (a) Additional GIWAXS patterns of Ba-Zr-S-O films synthesized at different temperatures. The measurements were performed with grazing incidence angles ($\alpha_i$) of 1.0° and 1.5°. Based on the positions of reflections in the patterns, three phases were identified: BaZrS$_3$[1], BaZrO$_3$[1] and ZrO$_2$[2]. Reference patterns of these phases are reported on the x-axis of the GIWAXS scans. (b) Crystal structure representations of BaZrS$_3$ and BaZrO$_3$. (c) Estimated penetration depth of X-rays for different grazing incidence angles used in this study.

**Compositional measurement of Ba-Zr-S-O thin film sulfurized at 1000 °C: RBS characterization**

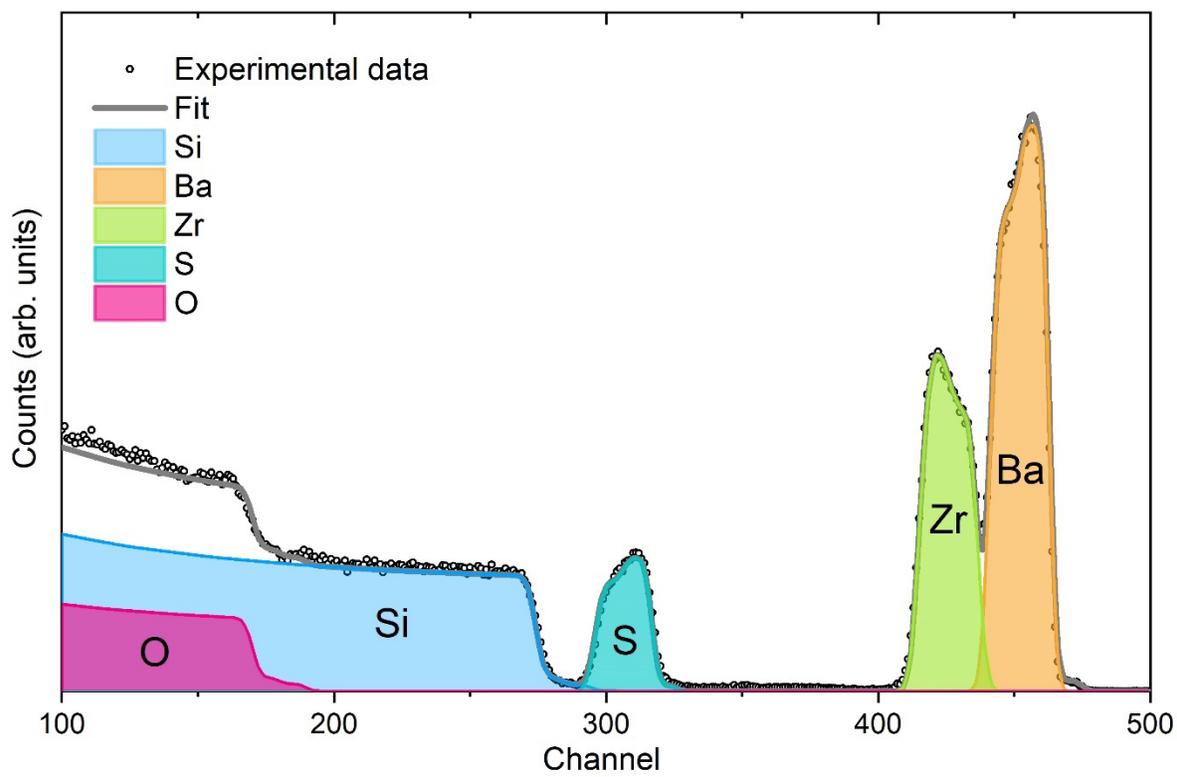

**Figure S7.** RBS spectra of Ba-Zr-O-S sample sulfurized at 1000 °C with fits corresponding to different element contributions, including Ba, Zr, S, O and Si.

**Details on RBS measurements and analysis**

**Rutherford Backscattering Spectrometry (RBS):**

RBS spectra are acquired at a backscattering angle of 160° and an appropriate grazing angle (with the sample oriented perpendicular to the incident ion beam). The sample is rotated or tilted with a small angle to present a random geometry to the incident beam. This avoids channeling in both the film and the substrate. The use of two detector angles can significantly improve the measurement accuracy for composition when thin surface layers need to be analyzed. The schematic diagram below shows the scattering geometry in a typical RBS experiment.

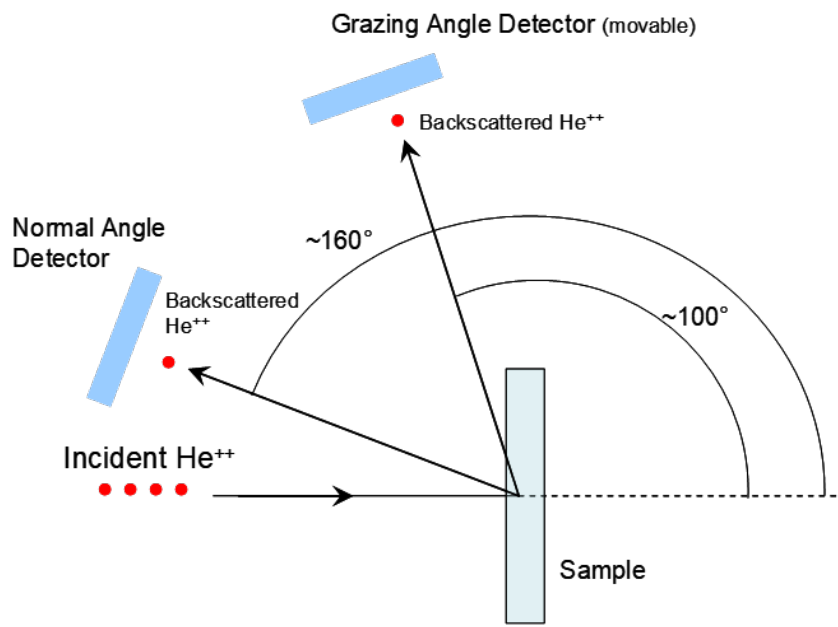

**Figure S8.** Scattering geometry of the RBS measurements

When a thin (<100nm) amorphous or polycrystalline film resides on a single crystal substrate "ion channeling" may be utilized to reduce the backscattering signal from the substrate. This results in improved accuracy in the composition of layers containing elements that overlay with the substrate signal, typically light elements such as oxygen, nitrogen and carbon.

| Analytical Parameters: RBS | |
|---|---|
| He++ Ion Beam Energy | 2.275MeV |
| Normal Detector Angle | 160° |
| Grazing Detector Angle | ~100° |
| Analysis Mode | *CC RR* |

Spectra are fit by applying a theoretical layer model and iteratively adjusting the concentrations and thickness until good agreement is found between the theoretical and the experimental spectra. These samples are modeled as a mixture of the film and substrate.